\documentclass[a4paper,twocolumn]{esapub} 

\usepackage{natbib, nopageno}
\usepackage[dvips]{graphicx}

\title{The SCUBA Local Universe Galaxy Survey}
\author[1]{Catherine Vlahakis}
\affil[1]{School of Physics and Astronomy, 
Cardiff University, 5 The Parade, Cardiff, CF24 3YB, United Kingdom, 
Email: Catherine.Vlahakis@astro.cf.ac.uk, Steve.Eales@astro.cf.ac.uk}
\author[2]{Loretta Dunne}
\affil[2]{School of Physics and Astronomy, 
University of Nottingham, Nottingham, NG7 2RD, United Kingdom, 
Email: Loretta.Dunne@nottingham.ac.uk}
\author[1]{Stephen Eales}

\begin{document}


\maketitle

\begin{abstract}
We present new results from the SCUBA Local Universe Galaxy Survey (SLUGS), the first large statistical submillimetre survey of the local Universe.
	Following our initial survey of a sample of 104 \textit{IRAS}-selected galaxies we now present the results of a sample of 80 Optically-Selected galaxies. This new sample, by definition free from temperature selection effects, allows us for the first time to determine how the amount of cold dust in galaxies varies with Hubble type. We detect 6 ellipticals in the sample and find them to have dust masses in excess of $10^{7}$ $M_{\odot}$. We derive local submillimetre Luminosity Functions and Dust Mass Functions, both directly for the Optically-Selected SLUGS and by extrapolation from the \textit{IRAS} PSCz survey, and find excellent agreement.
\end{abstract}

\section{Introduction}

Relatively little is known about the submillimetre properties of ``normal'' galaxies in the local universe -- \textit{IRAS} detected only the 10\% of dust warm enough to emit in the far-IR [2] and prior to SCUBA there existed only a handful of submillimetre flux measurements or maps.

\begin{figure}[hb]
\begin{centering}
\includegraphics[angle=270,width=1.7in]{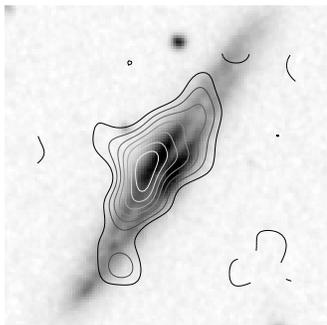}
\caption{Example from the optically-selected SLUGS: NGC 3987. 850 $\mu$m S/N map ($1\sigma$ contours) overlaid onto \textit{Digitised Sky Survey} optical image.}
\end{centering}
\end{figure}

The SCUBA Local Universe Galaxy Survey (SLUGS) is the first large systematic submillimetre survey of nearby galaxies. It consists of a sample selected from the \textit{IRAS} Bright Galaxy Sample [3], and a sample selected from the CfA optical redshift survey [6]. Since SCUBA is sensitive to the large proportion of dust too cold to be detected by \textit{IRAS}, whereas the \textit{IRAS}-selected sample is likely to have been biased toward galaxies containing warmer dust the Optically-Selected sample should, by definition, be free from temperature selection effects and would include any ``cold'' ($T_{dust}<$30K) galaxies present.

\section{The Optically-selected SLUGS: Results}

The 60, 100 and 850$\mu$m flux densities are well-fitted by single-temperature dust spectral energy distributions, with a mean best-fitting temperature for the sample $T_{dust}$=32.4K and a mean dust emissivity index $\beta$=0.97. This low value of $\beta$ may indicate that galaxies, across all Hubble types, contain a significant proportion of dust colder than these temperatures.

\begin{figure}[hb]
\begin{centering}
\includegraphics[angle=0,width=3.2in]{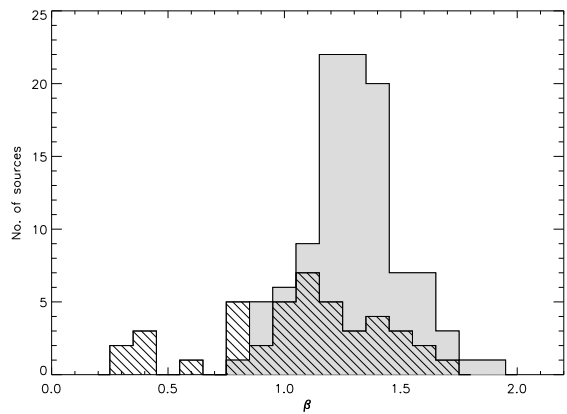}
\caption{Distribution of $\beta$ for Optically- and \textit{IRAS}-selected SLUGS (line-filled and shaded respectively).}
\end{centering}
\end{figure}

For the Optically-selected sample we see consistently lower values of $\beta$ than the \textit{IRAS} sample, strongly indicating the presence of a significantly larger fraction of cold dust than in the \textit{IRAS}-selected galaxies. Moreover, the marked differences between the distributions of $\beta$ and $T_{dust}$ for the two samples indicates that galaxies across all Hubble types contain significant amounts of cold dust.

This can be seen clearly by comparing the submm:far-IR colours for the two SLUGS samples -- we find a strong correlation between the 60:100$\mu$m and 60:850$\mu$m colours. The Optically-selected galaxies occupy a distinctly different region of the plot, with much colder colours. This, together with the marked differences between the distributions of $\beta$ and $T_{dust}$ for the two samples, indicates that not only are the two samples \textit{not} sampling the same populations of galaxies but also that the Optically-selected sample is probing a population of cold dusty galaxies ``missed'' by \textit{IRAS}.

\begin{figure}
\begin{centering}
\includegraphics[angle=0,width=3.1in]{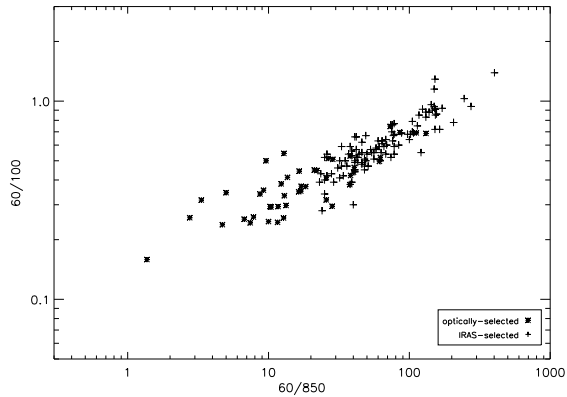}
\caption{\label{col}60:100$\mu$m versus 60:850$\mu$m Colour-Colour plot for the Optically- and \textit{IRAS}-selected SLUGS.}
\end{centering}
\end{figure}

\subsection[Ellipticals in the Optical SLUGS]
{Ellipticals in the Optical SLUGS}

It was once thought that ellipticals were entirely devoid of dust and gas, but optical absorption studies now show that dust is usually present.  Dust masses for the $\sim15\%$ of ellipticals detected by \textit{IRAS} have been found to be as much as a factor of 10--100 higher when estimated from their FIR emission compared to estimates from optical absorption, suggesting a diffuse cold dust component [4] (and refs. therein) and [1]. 
At  $850\mu$m we detect 6 ellipticals, from a total of 11 ellipticals in the optically-selected sample, and find them to have dust masses in excess of $10^{7}$ $M_{\odot}$. We are investigating this further with SCUBA observations of a larger sample of ellipticals.

\subsection[The $850\mu$m Luminosity Function]
{The $850\mu$m Luminosity Function}

\begin{figure}
\begin{centering}
\includegraphics[angle=270,width=3.0in]{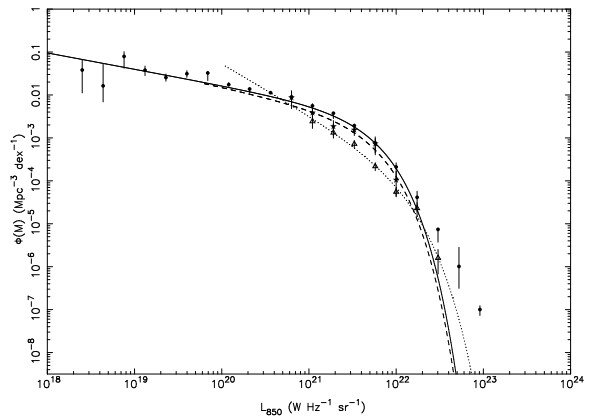}
\caption{\label{LF}PSCz-extrapolated $850\mu$m LF compared with those directly measured from Optically- and \textit{IRAS}-selected SLUGS samples (circles, stars, triangles respectively).}
\end{centering}
\end{figure}

Deep SCUBA surveys have been carried out, but a significant limitation in the interpretation of high-redshift surveys and studies of cosmological evolution has been the lack of a direct \textit{local} measurement of the sub-mm Luminosity Function. Until now, most deep sub-mm investigations have started out from a local \textit{IRAS} 60$\mu$m LF and extrapolated to the sub-mm, which as shown by [3] does not produce the measured sub-mm LF. Using the \textit{IRAS}-selected sample [3] produced the first direct estimate of the sub-mm LF. However, since the \textit{IRAS} sample is biased toward galaxies with larger amounts of warmer dust its LF may also be subject to bias. 

We have derived a direct $850\mu$m LF for the Optically-Selected sample. However, in order to better constrain the LF at the lower luminosity end we need more data points and we need to probe a wider range of luminosities. We do this using a method described by [5]: we determine an $850\mu$m LF using $\sim10000$ galaxies from the \textit{IRAS} PSCz survey -- we predict their $850\mu$m luminosities by extrapolating their \textit{IRAS} fluxes using our colour-colour relation (Fig.~\ref{col}). In Fig.~\ref{LF} we compare the PSCz-extrapolated and direct Optically-selected LFs. We find them to be well-fitted by Schechter functions and show that, whereas the slope of the \textit{IRAS}-selected LF at lower luminosities was steeper than -2 (a submm ``Olbers' Paradox''), the PSCz-extrapolated LF, as expected, flattens out at the low luminosity end and has a slope of -1.34.
We also compare the \textit{IRAS}-selected sample LF and find that such a sample consistently underestimates the 850$\mu$m LF. 



\end{document}